\documentclass[aps,pra,10pt,twocolumn,amsmath,amssymb,showpacs]{revtex4}
\usepackage{graphicx}
\renewcommand\Re{\operatorname{Re}}
\renewcommand\Im{\operatorname{Im}}
\renewcommand{\vec}[1]{\mathbf#1}
\newcommand{\vev}[1]{\langle#1\rangle}
\DeclareMathOperator{\tr}{tr}
\DeclareMathOperator{\Li}{Li}

\begin{document}

\title{Quantum critical transport in the unitary Fermi gas}
\author{Tilman Enss}

\affiliation{Physik Department, Technische Universit\"at M\"unchen,
  D-85747 Garching, Germany} 

\begin{abstract}
  The thermodynamic and transport properties of the unitary Fermi gas
  at finite temperature $T$ are governed by a quantum critical point
  at $T=0$ and zero density.  We compute the universal shear viscosity
  to entropy ratio $\eta/s$ in the high-temperature quantum critical
  regime $T\gg |\mu|$ and find that this strongly coupled quantum
  fluid comes close to perfect fluidity $\eta/s=\hbar/(4\pi k_B)$.
  Using a controlled large-$N$ expansion we show that already at the
  first non-trivial order the equation of state and the Tan contact
  density $C$ agree well with the most recent experimental
  measurements and theoretical Luttinger-Ward and Bold Diagrammatic
  Monte Carlo calculations.
\end{abstract}


\maketitle


\section{Introduction}
\label{sec:intro}

\begin{figure}[t]
  \centering
  \includegraphics[width=\linewidth]{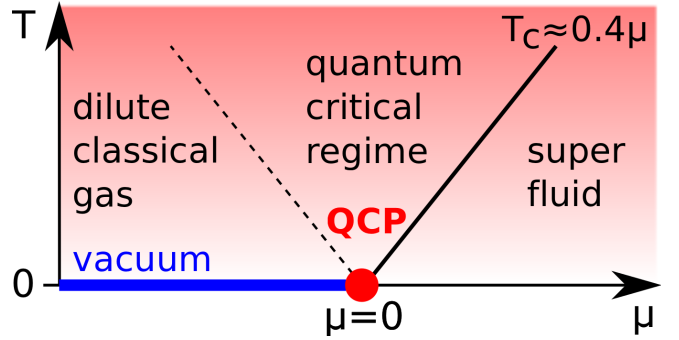}
  \caption{Universal phase diagram of the unitary Fermi gas.}
  \label{fig:phase}
\end{figure}

The unitary Fermi gas is a basic many-body problem which describes
strongly interacting fermions ranging from ultracold atoms near a
Feshbach resonance \cite{ketterle2008, bloch2008, giorgini2008} to
dilute neutron matter.  The properties in the dilute limit are
independent of the microscopic details of the interaction potential
and share a common universal phase diagram.  A quantum critical point
(QCP) at zero temperature governs the critical behavior in the whole
phase diagram as a function of temperature $T$, chemical potential
$\mu$, detuning from the Feshbach resonance $\nu$, and magnetic field
$h$ \cite{sachdev1999, nikolic2007, sachdev2012}.  Whereas
conventional QCPs separate two phases of finite density, in our case
the density itself is the order parameter which vanishes for $\mu<0$
and assumes a finite value for $\mu>0$ \cite{sachdev2012}.  In the
spin balanced case $h=0$, and at resonance $\nu=0$ the Fermi gas is
unitary and scale invariant.  In terms of the thermal length
$\lambda_T = \hbar(2\pi/mk_BT)^{1/2}$ the density equation of state
$n\lambda_T^3 = f_n(\mu/k_BT)$ is a universal function which has been
measured experimentally \cite{nascimbene2010, ku2012}.  The unitary
Fermi gas becomes superfluid at a universal $T_c(\mu) \approx
0.4\,\mu$ \cite{ku2012}, see Fig.~\ref{fig:phase}.  In this work we
focus on the quantum critical regime $T>0$ above the QCP at $h=0$,
$\nu=0$ and $\mu=0$, where $n\lambda_T^3 = f_n(0) \approx 2.9$ is a
universal constant.  Since the thermal length $\lambda_T$ is
comparable to the mean particle spacing $n^{-1/3}$, quantum and
thermal effects are equally important.  There is no small parameter,
and it is a theoretical challenge to compute the critical properties.
Recent measurements \cite{ku2012} and computations
\cite{vanhoucke2012a, drut2012} of the equation of state now agree to
the percent level.  However, a precise determination of transport
properties is much more demanding.

In order to reliably estimate transport coefficients we perform
controlled calculations in a large-$N$ expansion \cite{nikolic2007,
  veillette2007}.  Due to the lack of an intrinsic small parameter we
introduce an artificial small parameter, $1/N$, which organizes the
different diagrammatic contributions, or scattering processes, into
orders of $1/N$.  The original theory is recovered in the limit $N=1$.
One can perform controlled calculations by including all diagrams up
to a certain order in $1/N$, and these approximations can be
systematically improved by going to higher order.  This approach is
similar to the $\varepsilon$ expansion in the dimension of space.  The
advantage over perturbation theory is that it is controlled even at
strong interaction, while in contrast to Quantum Monte Carlo it works
directly in the thermodynamic limit and needs no finite size scaling.

We thus obtain new results for the Tan contact density
\cite{tan2008energetics, tan2008large, braaten2012} and the transport
properties in the quantum critical region.  The shear viscosity $\eta
= \hbar \lambda_T^{-3} f_\eta(\mu/k_BT)$ assumes a universal value at
$\mu=0$.  In kinetic theory $\eta=P\tau$ is given by the pressure $P$
times the viscous scattering time $\tau$, which is related to the
incoherent relaxation time of the gapless critical excitations above
the QCP.  The entropy density $s=k_B \lambda_T^{-3} f_s(\mu/k_BT)$ at
$\mu=0$ is exactly proportional to the pressure, $s=5P/2T$, and the
viscosity to entropy ratio (at $N=1$)
\begin{align}
  \label{eq:etas1}
  \frac{\eta}{s} 
  = \frac 25 T\tau
  \approx 0.74\,\frac{\hbar}{k_B}
\end{align}
is a universal number \emph{independent of temperature}.  A
temperature independent ratio $\eta/s=\hbar/(4\pi k_B)$ has been found
in certain string theories \cite{policastro2001} and is conjectured to
hold as a lower bound in other models \cite{kovtun2005}.  Strongly
interacting quantum fluids which saturate this bound are called
perfect fluids \cite{schaefer2009}.  Among real non-relativistic
fluids the unitary Fermi gas comes closest to the bound and is almost
perfect \cite{enss2011, cao2011, wlazlowski2012}, while for graphene
the viscosity decreases logarithmically with temperature in the
quantum critical regime \cite{mueller2009}.

We compare our large-$N$ results at $N=1$ \cite{nozieres1985} with
experimental measurements \cite{ku2012, cao2011, cao2011njp,
  kuhnle2011} and other theoretical approaches, including
self-consistent Luttinger-Ward \cite{haussmann1994, haussmann2007,
  enss2011} and Bold Diagrammatic Monte Carlo (BDMC)
\cite{vanhoucke2012a} calculations, see Table~\ref{tab:vals}.
\begin{table}[t]
  \centering
  \begin{tabular}{lllll}
    \hline
    & Experiment & Large-$N$ & LuttWard & BoldDiagMC \\
    \hline\hline
    $n \lambda_T^3$ & 2.966(35) \cite{ku2012} & 2.674 & 3.108
    \cite{haussmann2007} & 2.90(5)\;\: \cite{vanhoucke2012a} \\
    $P\;[nk_BT]$ & 0.891(19) \cite{ku2012} & 0.928 &
    0.863 \cite{haussmann2007} & 0.90(2)\;\: \cite{vanhoucke2012a} \\
    $s\;[nk_B]$ & 2.227(38) \cite{ku2012} & 2.320 & 2.177
    \cite{haussmann2007} & 2.25(5)\;\: \cite{vanhoucke2012a} \\
    $C\;[k_F^4]$ & & 0.0789 & 0.084 \cite{enss2011} &
    0.080(5) \cite{vanhoucke2012b} \\
    $\eta/s\;[\hbar/k_B]$ & 1.0(2) \cite{cao2011, schaefer2012} &
    0.741 & 0.708 \cite{enss2011} & \\
    \hline
  \end{tabular}
  \caption{Thermodynamic properties and transport coefficients of the
    unitary Fermi gas in the quantum critical region $\mu=0$, $T>0$:
    density $n$, pressure $P$, entropy density $s$, Tan contact
    density $C$, and shear viscosity $\eta$, with Fermi momentum $k_F
    = (3\pi^2n)^{1/3}$.  Large-$N$ results extrapolated to $N=1$.} 
  \label{tab:vals}
\end{table}

The excellent agreement between experiment and BDMC provides a
reliable reference to assess the accuracy of other methods.  We find
very good agreement of the pressure $P$ with large-$N$ ($3\%$ above
BDMC) and Luttinger-Ward ($4\%$ below) calculations, just slightly
outside the error bars, and we find similarly good agreement for the
entropy density $s$.  From the BDMC equation of state simulations of
\cite{vanhoucke2012a}, one can extract (via the pair propagator) a
preliminary value for the contact density \cite{vanhoucke2012b} $C /
k_F^4 = 0.080(5)$.  Our large-$N$ value is just $1.4\%$ below the BDMC
value, which is remarkable given how simple the calculation is, while
the Luttinger-Ward value lies about $5\%$ above the BDMC value, just
inside the error bars.  Experimental measurements of the contact
\cite{kuhnle2011} yield $C = 0.030(6)\, k_F^4$ for the trapped gas at
$\mu=0$ ($T/T_F=0.64$), which agrees well with trap averaged
calculations \cite{kuhnle2011}.  However, knowledge of the trap
averaged contact does not allow us to reconstruct the corresponding
value for the homogeneous system, so we refrain from a direct
comparison.  Dynamical and transport properties such as $\eta/s$ are
harder to compute than thermodynamic properties, which makes simple
approximations all the more valuable: we find that $\eta/s$ agrees to
$5\%$ between large-$N$ and Luttinger-Ward theory, giving a narrow
estimate.  The viscosity of a trapped gas has been measured
experimentally and agrees with trap averaged calculations
\cite{cao2011, cao2011njp, schaefer2012}, but differs from the
viscosity of the homogeneous system.

The body of this paper explains how these values are obtained: in
section~\ref{sec:ufg} we review the renormalization group (RG)
analysis of the unitary Fermi gas and its universal phase diagram, in
section~\ref{sec:largen} we perform thermodynamic and transport
calculations using the controlled large-$N$ expansion, and in
section~\ref{sec:LW} we extract the $\mu=0$ data from the
self-consistent Luttinger-Ward calculation, before concluding in
section~\ref{sec:disc}.  In particular, in Appendix~\ref{app:tan} we
give a new derivation of the Tan adiabatic and energy relations and
show that they are satisfied \emph{exactly} in self-consistent
Luttinger-Ward approximations, while Appendix~\ref{app:boltz} provides
technical details on the quantum kinetic equation.


\section{Phase diagram of the unitary Fermi gas}
\label{sec:ufg}

The interacting two-component Fermi gas is described by the action
\begin{multline}
  \label{eq:Sferm}
  S_F = \int d^dx\, d\tau\, \Bigl\{ \sum_\sigma \psi_{\sigma}^*
    \Bigl( \partial_\tau - \frac{\nabla^2}{2m} - \mu_\sigma \Bigr)
    \psi_{\sigma} \\
  + g_0 \psi_{\uparrow}^* \psi_{\downarrow}^* \psi_{\downarrow}
    \psi_{\uparrow} \Bigr\}
\end{multline}
where $\psi_\sigma$ are Grassmann variables representing fermion
species $\sigma = \;\uparrow, \downarrow$ of equal mass $m$, and the
imaginary time $\tau=0\dotsc\beta$ runs up to the inverse temperature
$\beta=1/T$ (we use units where $\hbar = 1 = k_B$).  $\mu_\sigma$ is
the chemical potential of species $\sigma$, but we will only
consider the spin-balanced case $\mu=\mu_\uparrow = \mu_\downarrow$.

In $d=3$ dimensions the scattering amplitude for small relative
momenta $k$ can be written in the form \cite{bloch2008}
\begin{align}
  f(k)=\frac{1}{-1/a-ik+r_e k^2/2}
\end{align}
where the scattering length $a$ can be varied experimentally by an
applied magnetic field, and the effective range $r_e$ depends on the
details of the interatomic potential.  By fine-tuning to a Feshbach
resonance $1/a\to 0$ the two-particle scattering remains strong at low
energy $k\to0$ and reaches the unitarity limit $f(k)=i/k$ independent
of $r_e$.  The low-energy properties remain universal at finite
density $n>0$ if $r_e$ is much shorter than the mean particle spacing
$n^{-1/3}$.  This condition $k_Fr_e\to 0$ is realized physically for a
dilute gas and near a broad Feshbach resonance as in $^6$Li
\cite{bloch2008}.

A finite $r_e$ regularizes the contact interaction
at short distances (UV), and for a sharp momentum cutoff $\Lambda \sim
1/|r_e|$ the detuning $\nu$ is related to the bare coupling $g_0$ in
\eqref{eq:Sferm} by
\begin{align}
  \nu \equiv -\frac 1a = -\frac{4\pi}{m} \left( \frac{1}{g_0} +
    \frac{m\Lambda}{2\pi^2} \right) \,.
\end{align}
Note that the resonance $\nu=0$ can only be reached for attractive
interactions $g_0<0$, when a bound state of the interatomic potential
is at the continuum threshold.

More generally, this can be understood from an RG analysis of the
model \eqref{eq:Sferm}: at zero temperature and density the running
coupling $g$ obeys the \emph{exact} flow equation \cite{sachdev1999,
  nikolic2007, sachdev2012}
\begin{align}
  \label{eq:beta}
  \frac{dg}{d\ell} = (2-d)g - \frac{g^2}{2}
\end{align}
which in $2<d<4$ has an unstable fixed point at $g_*=-2(d-2)<0$
corresponding to the Feshbach resonance.  For smaller $g<g_*$ the
fermions will form a BEC of fermion pairs; for larger $g>g_*$ the flow
runs toward the attractive fixed point $g=0$ of the free Fermi gas
(BCS limit).  At the Feshbach resonance fixed point the detuning $\nu$
is a relevant perturbation with scaling dimension $\dim[\nu]=d-2$.

The zero temperature phase diagram exhibits a quantum critical point
at the Feshbach resonance $\nu=0$, zero chemical potential
$\mu=0$, and zero spin imbalance $h=0$, where the ``magnetic
field'' $h$ couples to the difference in chemical potential
$\mu_\uparrow - \mu_\downarrow$.  This critical point determines a
universal phase diagram for finite $T$, $\nu$, $\mu$ and $h$
\cite{nikolic2007, sachdev2012}.  In this work we concentrate on the
spin balanced gas $h=0$ at unitarity $\nu=0$: the phase diagram for
finite $T$ and $\mu$ is depicted in Fig.~\ref{fig:phase}.

On the lower right for $\mu/T > (\mu/T)_c$ there is a superfluid phase
of fermion pairs, while the left part is a normal Fermi liquid phase
at finite density.  The phase transition line $T_c(\mu) \approx
0.4\mu$ \cite{ku2012} is universal and strictly linear, in contrast to
the corresponding phase diagram for a dilute Bose gas
\cite{sachdev1999}.  On the left for $\mu/T \to -\infty$ the Fermi
liquid crosses over to a dilute classical gas.  The line $T=0$,
$\mu<0$ has zero density (vacuum).  Here we focus on the
high-temperature quantum critical regime $T \gg |\mu|$, and in the
following we compute the thermodynamic and transport properties
specifically for the representative value $\mu=0$.

It is useful to perform a Hubbard-Stratonovich transformation to
decouple the fermion interaction.  We introduce a complex field
$\phi(x,\tau)$ representing a fermion pair and write the Bose-Fermi
action
\begin{multline}
  \label{eq:SBF}
  S_{BF} = \int d^dx\, d\tau\, \Bigl\{ \sum_\sigma \psi_{\sigma}^*
  \Bigl( \partial_\tau - \frac{\nabla^2}{2m} - \mu_\sigma \Bigr)
  \psi_{\sigma} \\
  - \frac{1}{g_0} |\phi|^2 - \phi \psi_\uparrow^* \psi_\downarrow^*
  - \phi^* \psi_\downarrow \psi_\uparrow \Bigr\} \,.
\end{multline}
Note that the pairing field $\phi$ has a positive gap because $g_0<0$
near the Feshbach resonance.  The action \eqref{eq:SBF} has the same
critical behavior as the two-channel atom-molecule model at its
zero-range fixed point \cite{nikolic2007, bloch2008}.

One can now proceed by integrating out the fermions to obtain an
effective bosonic action for the pairing field $\phi$.  This action
has bosonic vertices with any even number $2n$ of fields which are
given by a bare fermion loop with $2n$ vertex insertions.  In contrast
to the repulsive Fermi gas, where these vertices are irrelevant in the
RG sense, for the unitary Fermi gas these vertices all have marginal
scaling.  Already the particle-particle loop ($n=1$), which
contributes to the self-energy of the $\phi$ field, changes the bare
scaling dimension $\dim[\phi]=d/2$ of the $\phi$ field by an anomalous
contribution $\eta_\phi = 4-d$ to the true scaling dimension
$\dim[\phi]= (d+\eta_\phi)/2=2$, which is independent of $d$ (for
$2<d<4$).  Similarly, all higher bosonic vertices $n>1$ are singular
for small external frequencies and momenta and scale marginally in the
RG sense.  There is no small parameter to suppress these higher loop
diagrams, and they are \emph{a priori} equally important in the
infrared (IR).  At zero density the $2n>2$ particle $\beta$ functions
are decoupled from the $2n=2$ particle $\beta$ function in
Eq.~\eqref{eq:beta}, which is therefore exact.  Nevertheless, there
may also be a three-particle resonance (Efimov effect) in the
three-particle $\beta$ function depending on the mass ratio and
whether the particles are fermions or bosons \cite{nishida2008}.  This
changes the ground state from a two-particle to a three-particle bound
state and leads to limit cycles in the RG flow \cite{moroz2009}.

At finite density all higher bosonic vertices couple back into the
self-energy of the $\phi$ field.  In order to assess the quantitative
importance of these higher vertices, one can introduce an artificial
expansion parameter such as the dimension $\epsilon = 4-d$ for $2<d<4$
\cite{nishida2007fermi} or $1/N$ for a large number of fermion flavors
$N$ \cite{nikolic2007, veillette2007}.  Alternatively, one can use a
Monte Carlo sampling of diagrams \cite{vanhoucke2012a}.  In this work
we perform a large-$N$ expansion and compare it with the results from
other approaches.


\section{Large-N expansion}
\label{sec:largen}

We modify the Bose-Fermi action \eqref{eq:SBF} by introducing $N$
identical copies, or flavors, of $\uparrow$ and $\downarrow$ fermions,
denoted by $\psi_{\sigma a}$ with $\sigma=\;\uparrow,\downarrow$ and
the flavor index $a=1,\dotsc,N$.  The pairing field $\phi$ is chosen
to create an $\uparrow\downarrow$ pair of any flavor, and we obtain
the action \cite{veillette2007, sachdev2012}
\begin{multline}
  \label{eq:SBFN}
  S_{BF} = \int d^dx\, d\tau \Bigl\{ \sum_{\sigma a} \psi_{\sigma a}^*
  \Bigl( \partial_\tau - \frac{\nabla^2}{2m} - \mu_\sigma \Bigr)
  \psi_{\sigma a} \\
  - \frac{N}{g_0} |\phi|^2
  - \phi \sum_a \psi_{\uparrow a}^* \psi_{\downarrow a}^*
  - \phi^* \sum_a \psi_{\downarrow a} \psi_{\uparrow a} \Bigr\} \,.
\end{multline}
This action is $O(N)$ invariant under rotations in flavor space.  The
Gaussian integral over the fermion field yields the effective bosonic
action
\begin{multline}
  \label{eq:SBN}
  S_B = N \int d^dx\, d\tau \Bigl\{ -\tr_\sigma \ln \left[ 
    \begin{smallmatrix}
      \partial_\tau - \frac{\nabla^2}{2m} - \mu_\uparrow & -\phi \\
      -\phi^* & \partial_\tau + \frac{\nabla^2}{2m} + \mu_\downarrow
    \end{smallmatrix} \right] \\
  - \frac{1}{g_0} |\phi|^2 \Bigr\} \\
  = N T\sum_{\omega_m} \sum_{\vec{k}} \Bigl\{ \sum_\sigma \ln
  G_{0\sigma}(k,\omega_m) \\
  - \mathcal T^{-1}(k,\omega_m) |\phi(k,\omega_m)|^2 
  + \mathcal O(|\phi|^{\geq 4}) \Bigr\}
\end{multline}
with the trace running over the spin index $\sigma$.  The bare Fermi
propagator $G_{0\sigma}(k,\omega_m)$ is given by
\begin{align}
  \label{eq:Gbare}
  G_{0\sigma}^{-1}(k,\omega_m) & = -i\omega_m + \varepsilon_k - \mu_\sigma
\end{align}
with dispersion $\varepsilon_k = k^2/(2m)$, and the bosonic propagator
$-\mathcal T(k,\omega_m)$ is given by the regularized $T$-matrix in
medium,
\begin{align}
  \label{eq:Tbare}
  \mathcal T^{-1}(k,\omega_m) & = \frac{1}{g_0} + T\sum_{\epsilon_n} \int
  \frac{d^dp}{(2\pi)^d} \, G_{0\uparrow}(p,\epsilon_n) \notag \\
  & \qquad \times G_{0\downarrow}(\vec{k}-\vec{p},\omega_m-\epsilon_n)
  \,.
\end{align}
The number of flavors $N$ appears only as a global prefactor in the
action \eqref{eq:SBN}, hence a controlled loop expansion is possible
\cite{nikolic2007}. Each closed fermion loop contributes a factor of
$N$, while each $\phi$ propagator is suppressed by $1/N$. Even though
the higher bosonic vertices still have marginal scaling, their
contributions to the grand potential are now suppressed quantitatively
by powers of $1/N$. For $T\gg \mu$ the system is in the normal phase,
and the action \eqref{eq:SBN} has a saddle point at $\vev\phi = 0$. To
order $\mathcal O(1/N)$ the grand potential reads
\begin{align}
  \label{eq:OmegaN}
  \frac{\Omega}{N} = 
  T\sum_{\omega_m} \sum_{\vec{k}} \Bigl\{ 2 \ln G_{0}(k,\omega_m)
  - \frac{1}{N} \ln \mathcal T(k,\omega_m) \Bigr\}
\end{align}
Note that this order of the $1/N$ expansion extrapolated to $N=1$ is
exactly the Nozi\`eres--Schmitt-Rink (NSR) theory \cite{nozieres1985}.
The Matsubara frequency summation can be continued analytically to
real frequency,
\begin{multline}
  \label{eq:OmegaReal}
  \frac{\Omega}{N} = \sum_k \Bigl\{ -2T\ln[1+e^{-\beta(\varepsilon_k-\mu)}] \\
  - \frac{1}{N} \int_{-\infty}^\infty \frac{d\omega}{\pi} \, b(\omega)
  \, \delta(k,\omega,\mu,\nu) \Bigr\}
\end{multline}
with the scattering phase shift $\delta(k,\omega,\mu,\nu) = \Im \ln
\mathcal T(k,\omega,\mu,\nu)$ and the Bose function $b(\omega) =
[\exp(\beta\omega)-1]^{-1}$.  Specifically in $d=3$ the $T$-matrix
reads (in the spin-balanced case $h=0$)
\begin{align}
  \mathcal T^{-1}(k,\omega) & = -\frac{m\nu}{4\pi} -
  \frac{m^{3/2}}{4\pi}
  \sqrt{\frac{\varepsilon_k}{2}-\omega-i0-2\mu} \notag \\
  & \quad + \frac{m}{2\pi^2k} \int_0^\infty dp\,
  \frac{p}{1+e^{\beta(\varepsilon_p-\mu)}} \notag \\
  \label{eq:T0}
  & \qquad \times \ln \left[
    \frac{\omega+i0+2\mu-\varepsilon_p-\varepsilon_{k-p}} 
    {\omega+i0+2\mu-\varepsilon_p-\varepsilon_{k+p}} \right] \,.
\end{align}
The integral is convergent and readily evaluated numerically.

\subsection{Thermodynamics}

Using \eqref{eq:T0} we obtain for the pressure $P=-\Omega/L^d$
(equation of state) at $\mu=0$, $\nu=0$, $h=0$ and $T>0$:
\begin{align}
  \label{eq:pN}
  \frac{P}{N} & = -\frac{\Omega}{NL^d} = P^{(0)} + \frac 1N P^{(1)} +
  \dotsm \notag \\
  & = \Bigl( 1.734\, 400 + \frac 1N 0.747\, 561 \Bigr) T\lambda_T^{-3}
\end{align}
where
\begin{align*}
  P^{(0)} & = 2(1-2^{-3/2}) \zeta(5/2)\, T\lambda_T^{-3} \\
  P^{(1)} & = \int \frac{d^3k}{(2\pi)^3} \frac{d\omega}{\pi}
  b(\omega) \delta(k,\omega) \,.
\end{align*}
Since the unitary Fermi gas is scale invariant the internal energy
density $\varepsilon$ is proportional to the pressure
\cite{ho2004universal}
\begin{align}
  \label{eq:eN}
  \frac{\varepsilon}{N} = \frac{3P}{2N}
  = \Bigl( 2.601\, 600 + \frac 1N 1.121\, 341 \Bigr) T\lambda_T^{-3} \,.
\end{align}
Also the entropy density $s=\partial P/\partial T = (\varepsilon+P-\mu
n)/T$ at unitarity and $\mu=0$ is proportional to the pressure,
\begin{align}
  \label{eq:sN}
  \frac{s}{N} = \frac{5P}{2TN} = \Bigl( 4.335\, 999 + \frac 1N
  1.868\, 902 \Bigr) \lambda_T^{-3} \,.
\end{align}
The density at $\mu=0$ to order $\mathcal O(1/N)$ is
\begin{align}
  \label{eq:nN}
  \frac{n}{N} & = \frac{d(P/N)}{d\mu} = n^{(0)} + \frac 1N n^{(1)} +
  \dotsm \notag \\
  & = \Bigl( 1.530\, 294 + \frac 1N 1.143\, 936 \Bigr) \lambda_T^{-3}
\end{align}
where
\begin{align*}
  n^{(0)} & = 2(1-2^{-1/2}) \zeta(3/2)\, \lambda_T^{-3} \\
  n^{(1)} & = \int \frac{d^3k}{(2\pi)^3} \, \frac{d\omega}{\pi}\,
  b(\omega) \frac{d\delta(k,\omega)}{d\mu} \,.
\end{align*}
If this order of the $1/N$ expansion is evaluated at $N=1$ (NSR) we
obtain for the density
\begin{align}
  \label{eq:nNSR}
  n = 2.674\, 230\, \lambda_T^{-3} \qquad (N=1).
\end{align}
The ratio of thermal length to mean particle spacing, $\lambda_T
n^{1/3} \approx 1.388$, is of order unity, hence quantum and thermal
fluctuations are equally important in the high-temperature quantum
critical region.  The density determines the Fermi temperature
\begin{align}
  \label{eq:efN}
  k_BT_F & = \frac{k_F^2}{2m} = \frac{(3\pi^2n)^{2/3}}{2m}
\end{align}
which is useful to compare with data given in terms of the reduced
temperature
\begin{align}
  \label{eq:tN}
  \theta \equiv \frac{T}{T_F}
  = \Bigl( \frac{3\sqrt\pi}{8} n\lambda_T^3 \Bigr)^{-2/3}
  = 0.681\, 496 \quad (N=1).
\end{align}

Finally, the Tan contact density is defined as the total spectral
weight (density) of the pairing field \cite{tan2008energetics,
  tan2008large, braaten2012}
\begin{align}
  \label{eq:CN}
  C & = m^2 \vev{\phi^* \phi}
  = -\frac{m^2}{N} \int \frac{d^3k}{(2\pi)^3} \frac{d\omega}{\pi}
  b(\omega) \Im \mathcal T(k,\omega) \notag \\
  & = 26.840\, 128 \frac{\lambda_T^{-4}}{N} \,.
\end{align}
At $N=1$ the contact can be expressed in terms of $k_F$ using
Eq.~\eqref{eq:efN} which yields $C=0.0789\,k_F^4$.  This is equivalent
to the non-self-consistent $T$ matrix result \cite{palestini2010} and
agrees with the BDMC calculation within $1.4\%$ (see
Table~\ref{tab:vals}), but it differs from the result in
\cite{sachdev2012} by a factor of two.

Note that the Tan adiabatic theorem \cite{tan2008large}
\begin{align}
  \label{eq:TanAd}
  \frac{d(-P/N)}{d\nu} = \frac{C}{4\pi m}
\end{align}
is fulfilled exactly in the $1/N$ expansion: the change of the
pressure with detuning is
\begin{align}
  \frac{d(-P/N)}{d\nu}
  & = -\frac 1N \int \frac{d^3k}{(2\pi)^3} \frac{d\omega}{\pi}
  b(\omega) \frac{d\delta(k,\omega)}{d\nu} \notag \\
  \label{eq:TanAd2}
  & = -\frac{m}{4\pi N} \int \frac{d^3k}{(2\pi)^3} \frac{d\omega}{\pi}
  b(\omega) \Im \mathcal T(k,\omega)
\end{align}
because the change of scattering phase shift with detuning is
$d\delta(k,\omega)/d\nu = (m/4\pi)\Im \mathcal T(k,\omega)$, and using
Eq.~\eqref{eq:CN} we obtain \eqref{eq:TanAd}.

\subsection{Transport}

At $N=\infty$ the fermions are free: once a shear flow is excited in
the infinite system it will continue forever, and the dynamic shear
viscosity is
\begin{align}
  \label{eq:etainf}
  \eta(\omega) = \pi P \delta(\omega) \,.
\end{align}
The Drude weight is proportional to the pressure, in accordance with
the viscosity sum rule \cite{taylor2010, enss2011, braby2011}.  At
order $1/N$ the fermions acquire a self-energy correction by
scattering off pairing fluctuations, so for large $N$ the fermions are
almost free quasi-particles with lifetime $\mathcal O(N)$ and an
energy shift of the quasi-particle dispersion $\Re \Sigma \sim 1/N$.
In kinetic theory the dynamic viscosity becomes
\begin{align}
  \label{eq:etainfomega}
  \eta(\omega) = \frac{P\tau}{1+(\omega \tau)^2}
\end{align}
with the viscous scattering time $\tau = \mathcal O(N)$: the
$\delta(\omega)$ function in \eqref{eq:etainf} is broadened to a peak
of width $1/N$ and height $N$.  Note that the high-frequency tail
$\eta \sim C/15\pi\sqrt{m\omega}$ \cite{enss2011} is not seen in
kinetic theory \cite{braby2011}.

In order to compute transport properties for large $N$ it is justified
to use the quantum Boltzmann equation \cite{damle1997, sachdev1999}:
(i) the fermions propagate as free particles between collisions, up to
subleading corrections, (ii) the collision integral $\Im \Sigma \sim
1/N$ contains only particle-particle scattering described by the
medium $T$-matrix $\mathcal T(k,\omega)$ [Eq.~\eqref{eq:Tbare}]
because particle-hole scattering appears at higher orders, and (iii)
in addition to the collision (dynamic) term there is a shift of the
dispersion (kinetic) term $\Re \Sigma \sim 1/N$ of the same order.
However, it is only a subleading correction to the leading real term
$\Pi_{xy}/N\sim N^0$ (see below) and can be neglected.

Based on these considerations we arrive at the Boltzmann equation
\cite{damle1997, massignan2005}
\begin{align}
  \label{eq:qukin}
  \frac{\partial f}{\partial t}
  + \dot{\vec r} \cdot \frac{\partial f}{\partial\vec{r}}
  + \dot{\vec p} \cdot \frac{\partial f}{\partial\vec{p}}
  = -\frac 1N I[f]
\end{align}
for the distribution function $f(\vec p,\vec r,t)$, where $I[f]$ is
the collision integral.  For the shear viscosity we consider a
velocity field $\vec u = u_x(y) \hat{\vec x}$ with a small shear
gradient $\partial u_x/\partial y$, and the local equilibrium
distribution $f(\vec p) = f^0(\epsilon-\vec u\cdot \vec p)$ with
$\epsilon=p^2/2m$.  In the stationary limit the Boltzmann equation
\eqref{eq:qukin} becomes \cite{massignan2005}
\begin{align}
  \label{eq:qukin2}
  -\frac{\partial u_x}{\partial y} v_y p_x \frac{\partial
    f^0}{\partial \epsilon} = -\frac 1N I[f] \,.
\end{align}
The velocity gradient induces a momentum current density
\begin{align}
  \label{eq:Pixy}
  \Pi_{xy} = 2N \int \frac{d^3p}{(2\pi)^3} \, v_y p_x f(\vec p)
  = -\eta \frac{\partial u_x}{\partial y}
\end{align}
proportional to $\partial u_x/\partial y$, with the coefficient given
by the shear viscosity $\eta$.  We choose a deviation from the
equilibrium distribution, $f = f^0 + \delta f$ with $\delta f =
f^0(1-f^0) \varphi(\vec p)$ and $\varphi(\vec p) = v_y p_x/T$, such
that the momentum current density is
\begin{align}
  \label{eq:PixyP}
  \Pi_{xy} & = \frac{2N}{T} \int \frac{d^3p}{(2\pi)^3} \, v_y^2 p_x^2 f_p^0
  (1-f_p^0) = P \,.
\end{align}
This is equal to the pressure for free fermions ($N=\infty$) at
arbitrary temperature, as can be seen by integrating by parts.  We can
now replace $-\partial u_x/\partial y = P/\eta$ in \eqref{eq:qukin2}
and take moments of the Boltzmann equation by integrating both sides
with $2N\int d^3p/(2\pi)^3\, v_y p_x$.  The left-hand side becomes
\begin{align}
  \label{eq:qukin3lhs}
  \frac{2NP}{\eta T} \int \frac{d^3p}{(2\pi)^3}\, v_y^2 p_x^2 f_p^0
  (1-f_p^0)
  = \frac{P^2}{\eta}
\end{align}
while the right-hand side yields the collision integral
\cite{massignan2005, bruun2005, braby2011}
\begin{align}
  \label{eq:cxy}
  C_{xy} & = 2\int \frac{d^3p}{(2\pi)^3}\, v_y p_x I[\delta f] \notag \\
  & = \frac 2T \int \frac{d^3p}{(2\pi)^3} \, v_y p_x 
  \int \frac{d^3p_1}{(2\pi)^3} \, \int d\Omega \,
  \frac{d\sigma}{d\Omega}\, |\vec v - \vec v_1| \notag \\
  & \quad \times f_p^0 f_{p_1}^0 (1-f_{p'}^0) (1-f_{p_1'}^0) \notag \\
  & \quad \times \bigl[ \varphi(\vec p) + \varphi(\vec p_1)
  - \varphi(\vec p') - \varphi(\vec p_1') \bigr]
\end{align}
where fermions with incoming momenta $\vec p$, $\vec p_1$ scatter into
outgoing momenta $\vec p'$, $\vec p_1'$.  It will be convenient to
express these momenta in terms of the total momentum $\vec q=\vec
p+\vec p_1$ and the relative momenta $\vec k=(\vec p-\vec p_1)/2$
($\vec k'=(\vec p'-\vec p_1')/2$) of the incoming (outgoing)
particles, with $|\vec k'|=|\vec k|$ by energy conservation.  The
occupation numbers give the probability that the incoming states are
occupied, and the outgoing states are not.  The differential cross
section is given by the medium $T$-matrix
\begin{align}
  \label{eq:dsigma}
  \frac{d\sigma}{d\Omega} = \Bigl\lvert \frac{m}{4\pi}
  \mathcal T(\vec p+\vec p_1, \omega=\varepsilon_p+\varepsilon_{p_1}-2\mu)
  \Bigr\rvert^2 \,. 
\end{align}
In the vacuum limit the center-of-mass scattering depends only on the
relative momentum $k$,
\begin{align}
  \label{eq:dsigmavac}
  \frac{d\sigma}{d\Omega} = \frac{a^2}{1+a^2k^2} \qquad \text{(vacuum)}
\end{align}
but at finite density there is an additional dependence on the total
momentum $q$ in the medium $T$-matrix $\mathcal T(q,k) = \mathcal
T(q,\omega= 2\varepsilon_{q/2}+ 2\varepsilon_k-2\mu)$.  In relative
coordinates the shear term in Eq.~\eqref{eq:cxy} is
\cite{smith1989}
\begin{align}
  \label{eq:DeltaPhi}
  \varphi(\vec p) + \varphi(\vec p_1) - \varphi(\vec p') - \varphi(\vec p_1')
  = \frac{k_x k_y - k_x' k_y'}{mT/2} \,.
\end{align}
The collision integral then reads
\begin{align}
  C_{xy} & = \frac{2}{\pi mT} \int \!\! \frac{d^3q}{(2\pi)^3} \int \!\!
  \frac{d^3k}{(2\pi)^3} k k_x k_y \lvert \mathcal T(q,k)\rvert^2
  f_{\vec q/2+\vec k}^0 f_{\vec q/2-\vec k}^0 \notag \\
  & \quad \times \int \frac{d\Omega_{k'}}{4\pi} (k_xk_y
    - k_x'k_y') (1-f_{\vec q/2+\vec k'}^0) (1-f_{\vec q/2-\vec k'}^0)
    \notag \\
  & = \frac{1}{30\pi^5 mT} \int dq\, q^2 \int dk\, k^7
  |\mathcal T(q,k)|^2 \notag \\
  \label{eq:cxy2}
  & \quad \times \bigl[I_{\ell=0}^2(q,k) - I_{\ell=2}^2(q,k)\bigr]
\end{align}
with the $\ell$-wave angular average $I_\ell(q,k)$ over the Fermi
distribution functions derived analytically in
appendix~\ref{app:boltz} (the $d$-wave average $I_{\ell=2}^2(q,k)$
contributes only $0.2\%$ to the integral \eqref{eq:cxy2}).  Finally,
only two integrals over the radial momenta $q$ and $k$ have to be
performed.  In the dilute classical regime the collision
integral can be computed analytically,
\begin{align}
  \label{eq:cxyclass}
  C_{xy}^\text{cl} = \frac{32\sqrt 2 z^2T^2\lambda_T^{-3}}{15\pi}
\end{align}
with fugacity $z=\exp(\beta\mu)$, and in the same limit the pressure
is $P_\text{cl} = 2zT\lambda_T^{-3}N$.  The viscosity is then given by
\cite{massignan2005}
\begin{align}
  \label{eq:etaclass}
  \eta_\text{cl}
  = \frac{P_\text{cl}^2}{C_{xy}^\text{cl}}
  = \frac{15\pi\lambda_T^{-3}N^2}{8\sqrt 2}
  = 4.165\, 203\, \lambda_T^{-3}N^2 \,.
\end{align}
In the high-temperature quantum critical regime $T>0$, $\mu=0$ the
collision integral has to be computed with the full medium $T$-matrix
$\mathcal T(q,\omega)$ from Eq.~\eqref{eq:Tbare}, which is done
numerically and yields
\begin{align}
  \label{eq:cxycrit}
  C_{xy} = 0.935\, 683\, T^2\lambda_T^{-3}
\end{align}
and together with the pressure at leading order in $1/N$, $P = NP^{(0)}
= 1.734\, 400\, T\lambda_T^{-3}N$, we obtain in the quantum critical regime
\begin{align}
  \label{eq:etacrit}
  \eta = \frac{P^2}{C_{xy}} = 3.214\, 917\, \lambda_T^{-3}N^2 \,.
\end{align}
This value is about $20\%$ lower than in the dilute classical limit
\eqref{eq:etaclass}, which is mostly due to the reduced pressure,
while the effects of reduced density and increased medium scattering
almost cancel each other in $C_{xy}$.  With the viscous relaxation
time $\tau = P/C_{xy}$ and the entropy density $s=5P/(2T)$ we obtain
the universal viscosity to entropy ratio independent of temperature,
\begin{align}
  \label{eq:etascrit}
  \frac{\eta}{s} = \frac 25 T \tau = 0.741\, 448\, \frac{\hbar N}{k_B} \,.
\end{align}
A related computation of the viscosity using the medium $T$-matrix has
been performed for large attractive interaction $k_Fa=-11.8$ which
found $\eta = 2.3 \hbar n$ for $\mu=0$ at $T/T_F = 0.7$
\cite{bruun2005}, slightly larger than our value \eqref{eq:etacrit} at
$N=1$.  Note that we have evaluated $\eta$ using only a single moment
of the Boltzmann equation \eqref{eq:qukin2}, but is has been shown
that corrections to $\eta$ from higher moments are less than $2\%$
\cite{bruun2007}.  A similar transport calculation using the medium
$T$-matrix in two dimensions has been performed recently
\cite{enss2012visc}.


\section{Luttinger-Ward theory}
\label{sec:LW}

The Luttinger-Ward theory provides a systematic way to obtain
self-consistent and conserving approximations, such that the Green's
functions satisfy all symmetries and conservation laws of the model
\cite{luttinger1960, baym1961}.  The Luttinger-Ward functional
$\Phi[G_\sigma,G_B]$ can be defined in terms of full fermionic
propagators $G_\sigma$ and full bosonic propagators $G_B = -\mathcal
T$.  The exact theory is given by an infinite set of irreducible
contributions to the $\Phi$ functional which cannot be evaluated in
practice, so typically one chooses a subclass of diagrams.  For the
unitary Fermi gas a very successful approximation is to use ladder
diagrams with full fermionic Green's functions \cite{haussmann1994,
  haussmann2007}.  Then the full $T$-matrix is given by an expression
similar to \eqref{eq:Tbare} but with full Green's functions,
\begin{multline}
  \label{eq:fullT}
  \mathcal T^{-1}(k,\omega_m) = \frac{1}{g_0} + T\sum_{\epsilon_n} \int
  \frac{d^3p}{(2\pi)^3}\, G_\uparrow(\vec p,\epsilon_n)\\
  \times G_\downarrow(\vec k-\vec p,\omega_m-\epsilon_n) \,.
\end{multline}
Since we are interested in the high-temperature critical region we
consider only the expressions valid in the normal phase.  The
Luttinger-Ward theory then prescribes that the $\uparrow$ fermionic
self-energy is given by scattering a $\downarrow$ fermion off pair
fluctuations described by the full $T$-matrix,
\begin{align}
  \label{eq:fullS}
  \Sigma_\uparrow(k,\omega_m) = T\sum_{\epsilon_n} \int
  \frac{d^3p}{(2\pi)^3}\, G_\downarrow(\vec p,\epsilon_n)\,
  \mathcal T(\vec k+\vec p,\omega_m+\epsilon_n)
\end{align}
and analogously for $\Sigma_\downarrow$.  The Dyson equation
determines the full fermionic Green's functions
\begin{align}
  \label{eq:fullG}
  G_\sigma^{-1}(k,\omega_m) = -i\omega_m + \varepsilon_k - \mu_\sigma -
  \Sigma_\sigma(k,\omega_m) \,.
\end{align}
This set of equations \eqref{eq:fullT}--\eqref{eq:fullG} is solved
self-consistently by iteration \cite{haussmann1994, haussmann2007}.
The resulting Green's functions in Matsubara frequency can be continued
analytically to obtain the spectral functions in real frequency, which
show substantial broadening near $T_c$ and additional excitations
beyond a single quasi-particle peak \cite{haussmann2009, gaebler2010,
  magierski2011}.  Similar features are observed in the spin polarized
case \cite{schmidt2011, kohstall2012, schmidt2012, koschorreck2012}.

The pressure $P=-\Omega/L^d$ is obtained from the grand potential
\cite{haussmann2007}
\begin{multline}
  \label{eq:fullOmega}
  \Omega = T\sum_{\omega_m} \sum_{\vec{k}} \Bigl\{ \sum_\sigma \ln
  G_\sigma(k,\omega_m) \\
  + \sum_\sigma [1-G_{0\sigma}^{-1}(k,\omega_m)
  G_\sigma(k,\omega_m)] - \ln \mathcal T(k,\omega_m) \Bigr\}
\end{multline}
evaluated using the self-consistent fermion propagator and the full
$T$-matrix.  We extract the high-temperature quantum critical behavior
from the existing thermodynamic data \cite{haussmann2007} interpolated
at $\mu=0$.  Specifically, we make a cubic spline interpolation of
$\mu(T)$ and find the solution of $\mu(T)=0$ at $\theta = T/T_F =
0.6165$, which implies $n\lambda_T^3 = 8/(3\sqrt\pi)\theta^{-3/2} =
3.108$.  Furthermore, we find $P = 0.8630\, nk_BT$, $s = 2.177\,
nk_B$, and $C = 0.084\, 353\, k_F^4$, which can be recast in terms of
$\lambda_T$.  These values are summarized in Table~\ref{tab:vals} and
are remarkably close to the experimental values.

The shear viscosity $\eta(T,\omega)$ has been computed in
Luttinger-Ward theory as a function of temperature and frequency
\cite{enss2011}: it has a Lorentzian peak at low frequency, followed
by a universal tail $\eta(T,\omega) \sim C(T)/15\pi\sqrt{m\omega}$
proportional to the contact density.  We make a cubic spline
interpolation of $\mu(\eta)=0$ and find the root at $\eta(T,\omega=0)
= 1.5409\, \hbar n$, which yields $\eta/s = 0.7077\, \hbar/k_B$.  This
result is slightly lower than the large-$N$ value in
Eq.~\eqref{eq:etascrit}.  We note that in this self-consistent
calculation the minimum of $\eta/s \approx 0.6 \hbar/k_B$ is found at
a somewhat lower temperature $T/T_F \approx 0.4$ \cite{enss2011}.


\section{Conclusions}
\label{sec:disc}

The unitary Fermi gas in the high-temperature quantum critical region
is a challenging many-body problem.  It is strongly interacting, with
the density almost twice the non-interacting value at $\mu=0$
\cite{ku2012}, and has no small expansion parameter.  Still, our
large-$N$ results at the first non-trivial order beyond the free Fermi
gas are already remarkably close to reliable experimental and
theoretical results \cite{ku2012, vanhoucke2012a}.  A main result of
the present paper is that this is true also for the transport
properties $\eta/s$ once medium effects are included in the quantum
kinetic equation.  A possible reason for this good agreement is that
large-$N$ and Luttinger-Ward approximations satisfy the Tan adiabatic
and energy relations exactly, as we show in Appendix~\ref{app:tan}.
In addition, Luttinger-Ward theory exactly fulfills the scale
invariance of the unitary Fermi gas \cite{enss2011}.  For a better
comparison between calculations for the homogeneous system and
experiments it would be desirable to have local measurements in the
spirit of Ref.~\cite{ku2012} also for the contact and transport
properties, since the comparison of trap averaged quantities is less
sensitive to the details of the temperature dependence.  A promising
step in this direction is to selectively probe atoms near the center
of the trap in order to extract the contact density from the tail of
the momentum distribution \cite{drake2012}.

\begin{acknowledgments}
  I wish to thank Lars Fritz, Subir Sachdev, J\"org Schmalian, Richard
  Schmidt, and Wilhelm Zwerger for fruitful discussions and Mark Ku,
  Thomas Sch\"afer, Chris Vale, F\'elix Werner, and Martin Zwierlein
  for sharing their data.
\end{acknowledgments}

\appendix


\section{Exact Tan relations in Luttinger-Ward theory}
\label{app:tan}

Consider the fermionic action \eqref{eq:Sferm}: a small variation of
the quadratic term, $\delta G_0^{-1}$, will lead to a change in the
grand potential
\begin{align}
  \label{eq:dOmega}
  \delta \Omega = -\tr(G \delta G_0^{-1})
\end{align}
with the trace running over space, time and possibly spin indices.
However, this equation is often violated if approximations are made
for the full Green's function $G$.  A unique feature of conserving
approximations, which are derived from a Luttinger-Ward functional
$\Phi[G]$, is that Eq.~\eqref{eq:dOmega} holds exactly even for
approximate $\Omega$ and $G$ \cite{baym1962}.

For the strongly interacting Fermi gas it is convenient to start from
the Bose-Fermi action \eqref{eq:SBF} and define a Luttinger-Ward
functional $\Phi[G_\sigma,G_B]$ in terms of both fermionic and bosonic
Green's functions \cite{haussmann1994, haussmann2007, enss2011}.  Then a
variation of the microscopic parameters $\delta G_{0\sigma}^{-1}$ and/or
$\delta G_{0B}^{-1}$ induces a change of the grand potential
\cite{enss2011}
\begin{align}
  \label{eq:dOmega2}
  \delta \Omega = -\tr(G_\sigma \delta G_{0\sigma}^{-1})
  + \tr(G_B \delta G_{0B}^{-1}) \,.
\end{align}
Again, this exact equation continues to hold within conserving
approximations with full self-consistent propagators $G_\sigma$ and
$G_B$.

We will now show that the Tan adiabatic theorem \cite{tan2008large}
\begin{align}
  \label{eq:tanad}
  \frac{d\Omega/L^d}{d(-1/a)} = \frac{C}{4\pi m}
\end{align}
and the Tan energy formula \cite{tan2008energetics}
\begin{align}
  \label{eq:tanen}
  \varepsilon = \sum_\sigma \int \frac{d^3k}{(2\pi)^3} \, \varepsilon_k
  \Bigl( n_{k\sigma} - \frac{C}{k^4} \Bigr) + \frac{C}{4\pi ma}
\end{align}
are consequences of \eqref{eq:dOmega2} and therefore hold not only in
the exact theory but in any conserving approximation, including the
self-consistent $T$-matrix approximation introduced in
section~\ref{sec:LW}.  A variation of detuning changes only the
bosonic quadratic term
\begin{align}
  \label{eq:G0B}
  G_{0B}^{-1}(k,\omega_m) = -\frac{1}{g_0}
  = \frac{m}{4\pi} \Bigl( -\frac 1a+\frac{2\Lambda}{\pi} \Bigr)
\end{align}
in the action \eqref{eq:SBF},
\begin{align}
  \label{eq:dG0ad}
  \frac{\partial G_{0\sigma}^{-1}(k,\omega_m)}{\partial(-1/a)} & = 0 &
  \frac{\partial G_{0B}^{-1}(k,\omega_m)}{\partial(-1/a)} & =
  \frac{m}{4\pi} \,.
\end{align}
The variation of the grand potential is then
\begin{align}
  \label{eq:dOmegaad}
  \frac{d\Omega}{d(-1/a)}
  & = -\tr\Bigl(G_\sigma \frac{\partial G_{0\sigma}^{-1}}
  {\partial(-1/a)}\Bigr)
  +\tr\Bigl(G_B \frac{\partial G_{0B}^{-1}}
  {\partial(-1/a)}\Bigr) \notag \\
  & = \frac{m}{4\pi} \tr(G_B)
\end{align}
with the density of bosons expressed by the Tan contact density,
\begin{multline}
  \label{eq:trGB}
  L^{-3} \tr(G_B)
  = T\sum_{\omega_m} \int \frac{d^3k}{(2\pi)^3} G_B(k,\omega_m)
  e^{+i0\omega_m} \\
  = G_B(x=0,\tau=-0)
  = \vev{\phi^* \phi}
  = \frac{C}{m^2} \,.
\end{multline}
Inserting \eqref{eq:trGB} into \eqref{eq:dOmegaad} directly yields the
adiabatic theorem \eqref{eq:tanad}.  In order to derive the energy
formula we consider a variation of mass,
\begin{align}
  \label{eq:inten}
  \varepsilon = m^{-1} \frac{d\Omega/L^3}{d(m^{-1})} \,.
\end{align}
Usually this yields only the kinetic energy (cf.\ Eq.~(61) in
\cite{baym1962}), but in our case the interaction term $4\pi a/m$ also
depends on mass, so \eqref{eq:inten} is the full internal energy
$\varepsilon=\vev H$ including the potential term.  Specifically,
\begin{align}
  \label{eq:dG0en}
  m^{-1}\frac{\partial G_{0\sigma}^{-1}(k,\omega_m)}{\partial m^{-1}}
  & = \varepsilon_k \\
  m^{-1}\frac{\partial G_{0B}^{-1}(k,\omega_m)}{\partial m^{-1}}
  & = \frac{m}{4\pi}  \Bigl( \frac 1a-\frac{2\Lambda}{\pi} \Bigr) \,,
\end{align}
and with the momentum distribution function $-T\sum_{\omega_m}
G_\sigma(k,\omega_m) = n_{k\sigma}$ we obtain the internal energy
density
\begin{align}
  \label{eq:dOmegaen}
  \varepsilon
  & = -\frac{m^{-1}}{L^3} \tr\Bigl(G_\sigma \frac{\partial G_{0\sigma}^{-1}}
  {\partial m^{-1}}\Bigr)
  +\frac{m^{-1}}{L^3} \tr\Bigl(G_B \frac{\partial G_{0B}^{-1}}
  {\partial m^{-1}}\Bigr) \notag \\
  & = \sum_\sigma \int^\Lambda \frac{d^3k}{(2\pi)^3} \varepsilon_k
  n_{k\sigma} + \frac{m}{4\pi} \Bigl( \frac 1a-\frac{2\Lambda}{\pi} \Bigr)
  \frac{C}{m^2} 
\end{align}
where the $k$ integral extends to the momentum cutoff $\Lambda$.  The
regularization term $\Lambda C/(2\pi^2m)$ can be written as
$\sum_\sigma \int d^3k/(2\pi)^3 \varepsilon_k C/k^4$, and we arrive at
the energy formula \eqref{eq:tanen}.  In a similar way the Tan
pressure relation has been derived in the Luttinger-Ward theory by an
infinitesimal scale transformation on the grand potential
\cite{enss2011}.  This concludes our proof that the Tan relations are
fulfilled exactly in the self-consistent $T$-matrix approximation.


\section{Quantum kinetic equation}
\label{app:boltz}

A useful feature of $d=3$ dimensions is that the angular averages of
the distribution functions can be performed analytically.  One can
write the product of Fermi functions in \eqref{eq:cxy2} with
$|\vec k'|=|\vec k|$ as
\begin{multline}
  \label{eq:distavg}
  f(\varepsilon_{\vec q/2+\vec k}) f(\varepsilon_{\vec q/2-\vec k})
  [1-f(\varepsilon_{\vec q/2+\vec k'})] [1-f(\varepsilon_{\vec
    q/2-\vec k'})] \\
  = \frac{1}{4(\cosh{a}+\cosh{bx})(\cosh{a}+\cosh{bx'})}
\end{multline}
with $x = \Hat{\vec k}\cdot \Hat{\vec q}$, $x' = \Hat{\vec k}'\cdot
\Hat{\vec q}$ and $a=(\varepsilon_{\vec q/2}+\varepsilon_{\vec
  k}-\mu)/T$, $b=kq/(2mT)$.  The angular average over the solid angles
of the vectors $\vec q$, $\vec k$ and $\vec k'$ is then
\begin{multline}
  \label{eq:angavg}
  \int \frac{d\Omega_{\vec q}}{4\pi}
  \int \frac{d\Omega_{\vec k}}{4\pi}
  \int \frac{d\Omega_{\vec k'}}{4\pi} k_x k_y (k_x k_y - k_x' k_y')
  ff[1-f][1-f] \\
  = \frac{k^4}{15} \Bigl( I_{\ell=0}^2(q,k) - I_{\ell=2}^2(q,k) \Bigr) 
\end{multline}
where we have defined the $\ell$-wave angular average of the
distribution functions
\begin{align}
  \label{eq:I02}
  I_\ell(q,k) = \frac 14 \int_{-1}^1 dx
  \frac{P_\ell(x)}{\cosh a + \cosh bx}
\end{align}
with Legendre polynomials $P_\ell(x)$.  The $s$-wave average is given
by
\begin{align}
  \label{eq:I0}
  I_{\ell=0}(q,k) = \frac{1}{2b\sinh a}
  \ln\frac{\cosh[(a+b)/2]} {\cosh[(a-b)/2]}
\end{align}
while the $d$-wave average can be expressed in terms of polylogarithms
$\Li_s(z)$,
\begin{multline}
  \label{eq:I2}
  I_{\ell=2}(q,k) = I_{\ell=0}(q,k)-\frac{1}{4b^3\sinh a} \\
  \times \Bigl[ 6\Li_3(-e^{a+b})-6\Li_3(-e^{b-a}) 
  -6b\Li_2(-e^{a+b}) \\ +6b\Li_2(-e^{b-a}) 
  + a(a^2-3ab+\pi^2) \Bigr].
\end{multline}
Thus, all angular integrations can be done analytically and only the
two radial integrations over $q$ and $k$ in Eq.~\eqref{eq:cxy2} need
to be performed numerically.


\begin{thebibliography}{43}

\bibitem{ketterle2008}
W.~Ketterle and M.~Zwierlein, Rivista del Nuovo Cimento \textbf{31}, 247–422
  (2008).

\bibitem{bloch2008}
I.~Bloch, J.~Dalibard, and W.~Zwerger, Rev.\ Mod.\ Phys.\ \textbf{80}, 885
  (2008).

\bibitem{giorgini2008}
S.~Giorgini, L.~P. Pitaevskii, and S.~Stringari, Rev.\ Mod.\ Phys.\ \textbf{80},
  1215 (2008).

\bibitem{sachdev1999}
S.~Sachdev, \emph{{Quantum Phase Transitions}} (Cambridge University Press,
  Cambridge, 1999).

\bibitem{nikolic2007}
P.~Nikoli{\'c} and S.~Sachdev, Phys.\ Rev.~A \textbf{75}, 033608 (2007).

\bibitem{sachdev2012}
S.~Sachdev, in \emph{{The BCS-BEC Crossover and the Unitary Fermi Gas}}, edited
  by W.~Zwerger (Springer, Berlin, 2012).

\bibitem{nascimbene2010}
S.~Nascimb{\`e}ne, N.~Navon, K.~J. Jiang, F.~Chevy, and C.~Salomon, Nature
  \textbf{463}, 1057 (2010).

\bibitem{ku2012}
M.~J.~H. Ku, A.~T. Sommer, L.~W. Cheuk, and M.~W. Zwierlein, Science
  \textbf{335}, 563 (2012).

\bibitem{vanhoucke2012a}
K.~Van~Houcke, F.~Werner, E.~Kozik, N.~Prokof'ev, B.~Svistunov, M.~J.~H. Ku,
  A.~T. Sommer, L.~W. Cheuk, A.~Schirotzek, and M.~W. Zwierlein (2012),
  Nature Phys.\ \textbf{8}, 366 (2012).

\bibitem{drut2012}
  J.~E. Drut, T.~A. L\"ahde, G.~Wlaz{\l}owski, and P.~Magierski,
  Phys.\ Rev.\ A \textbf{85}, 051601(R) (2012).

\bibitem{veillette2007}
M.~Y. Veillette, D.~E. Sheehy, and L.~Radzihovsky, Phys.\ Rev.~A \textbf{75},
  043614 (2007).

\bibitem{tan2008energetics}
S.~Tan, Ann.\ Phys.\ (N.Y.) \textbf{323}, 2952 (2008).

\bibitem{tan2008large}
S.~Tan, Ann.\ Phys.\ (N.Y.) \textbf{323}, 2971 (2008).

\bibitem{braaten2012}
E.~Braaten, in \emph{{The BCS-BEC Crossover and the Unitary Fermi Gas}}, edited
  by W.~Zwerger (Springer, Berlin, 2012).

\bibitem{policastro2001}
G.~Policastro, D.~T. Son, and A.~O. Starinets, Phys.\ Rev.\ Lett.\ \textbf{87},
  081601 (2001).

\bibitem{kovtun2005}
P.~K. Kovtun, D.~T. Son, and A.~O. Starinets, Phys.\ Rev.\ Lett.\ \textbf{94},
  111601 (2005).

\bibitem{schaefer2009}
T.~Sch{\"a}fer and D.~Teaney, Rep.\ Prog.\ Phys.\ \textbf{72}, 126001 (2009).

\bibitem{enss2011}
T.~Enss, R.~Haussmann, and W.~Zwerger, Ann.\ Phys.\ (N.Y.) \textbf{326}, 770
  (2011).

\bibitem{cao2011}
C.~Cao, E.~Elliott, J.~Joseph, H.~Wu, J.~Petricka, T.~Sch{\"a}fer, and J.~E.
  Thomas, Science \textbf{331}, 58 (2011).

\bibitem{wlazlowski2012}
G.~Wlaz{\l}owski, P.~Magierski, and J.~E. Drut, Phys.\ Rev.\ Lett.\
  \textbf{109}, 020406 (2012).

\bibitem{mueller2009}
M.~M{\"u}ller, J.~Schmalian, and L.~Fritz, Phys.\ Rev.\ Lett.\ \textbf{103},
  025301 (2009).

\bibitem{nozieres1985}
P.~Nozi{\`e}res and S.~Schmitt-Rink, J. Low Temp.\ Phys.\ \textbf{59}, 195
  (1985).

\bibitem{cao2011njp}
C.~Cao, E.~Elliott, H.~Wu, and J.~E. Thomas, New J. Phys.\ \textbf{13}, 075007
  (2011).

\bibitem{kuhnle2011}
E.~D. Kuhnle, S.~Hoinka, P.~Dyke, H.~Hu, P.~Hannaford, and C.~J. Vale, Phys.\
  Rev.\ Lett.\ \textbf{106}, 170402 (2011).

\bibitem{haussmann1994}
R.~Haussmann, Phys.\ Rev.~B \textbf{49}, 12975 (1994).

\bibitem{haussmann2007}
R.~Haussmann, W.~Rantner, S.~Cerrito, and W.~Zwerger, Phys.\ Rev.~A
  \textbf{75}, 023610 (2007).

\bibitem{vanhoucke2012b}
K.~Van~Houcke, F.~Werner, E.~Kozik, N.~Prokof'ev, and B.~Svistunov, private
  communication.

\bibitem{schaefer2012}
T.~Sch{\"a}fer and C.~Chafin, in \emph{{The BCS-BEC Crossover and the Unitary
  Fermi Gas}}, edited by W.~Zwerger (Springer, Berlin, 2012).

\bibitem{nishida2008}
Y.~Nishida, D.~T. Son, and S.~Tan, Phys.\ Rev.\ Lett.\ \textbf{100}, 090405
  (2008).

\bibitem{moroz2009}
S.~Moroz, S.~Floerchinger, R.~Schmidt, and C.~Wetterich, Phys.\ Rev.~A
  \textbf{79}, 042705 (2009).

\bibitem{nishida2007fermi}
Y.~Nishida and D.~T. Son, Phys.\ Rev.~A \textbf{75}, 063617 (2007).

\bibitem{ho2004universal}
T.~L. Ho, Phys.\ Rev.\ Lett.\ \textbf{92}, 090402 (2004).

\bibitem{palestini2010}
F. Palestini, A. Perali, P.~Pieri, and G.~C. Strinati,
  Phys.\ Rev.~A \textbf{82}, 021605(R) (2010).

\bibitem{taylor2010}
E.~Taylor and M.~Randeria, Phys.\ Rev.~A \textbf{81}, 053610 (2010).

\bibitem{braby2011}
M.~Braby, J.~Chao, and T.~Sch{\"a}fer, New J. Phys.\ \textbf{13}, 035014 (2011).

\bibitem{damle1997}
K.~Damle and S.~Sachdev, Phys.\ Rev.~B \textbf{56}, 8714 (1997).

\bibitem{massignan2005}
P.~Massignan, G.~M. Bruun, and H.~Smith, Phys.\ Rev.~A \textbf{71}, 033607
  (2005).

\bibitem{bruun2005}
G.~M. Bruun and H.~Smith, Phys.\ Rev.~A \textbf{72}, 043605 (2005).

\bibitem{smith1989}
H.~Smith and H.~H. Jensen, \emph{Transport Phenomena} (Oxford University
  Press, 1989).

\bibitem{bruun2007}
G.~M. Bruun and H.~Smith, Phys.\ Rev.~A \textbf{75}, 043612 (2007).

\bibitem{enss2012visc}
T.~Enss, C.~K{\"u}ppersbusch, and L.~Fritz, Phys.\ Rev.~A \textbf{86},
  013617 (2012).

\bibitem{luttinger1960}
J.~M. Luttinger and J.~Ward, Phys.\ Rev.\ \textbf{118}, 1417 (1960).

\bibitem{baym1961}
G.~Baym and L.~P. Kadanoff, Phys.\ Rev.\ \textbf{124}, 287 (1961).

\bibitem{haussmann2009}
R.~Haussmann, M.~Punk, and W.~Zwerger, Phys.\ Rev.~A \textbf{80}, 063612
  (2009).

\bibitem{gaebler2010}
J.~P. Gaebler, J.~T. Stewart, T.~E. Drake, D.~S. Jin, A.~Perali, P.~Pieri, and
  G.~C. Strinati, Nature Phys.\ \textbf{6}, 569 (2010).

\bibitem{magierski2011}
P.~Magierski, G.~Wlaz{\l}owski, and A.~Bulgac, Phys.\ Rev.\ Lett.\
  \textbf{107}, 145304 (2011).

\bibitem{schmidt2011}
R.~Schmidt and T.~Enss, Phys.\ Rev.~A \textbf{83}, 063620 (2011).

\bibitem{kohstall2012}
C.~Kohstall, M.~Zaccanti, M.~Jag, A.~Trenkwalder, P.~Massignan,
  G.~M. Bruun, F.~Schreck, and R.~Grimm, DOI:10.1038/nature11065 [Nature
  (to be published)].

\bibitem{schmidt2012}
R.~Schmidt, T.~Enss, V.~Pietil{\"a}, and E.~Demler, Phys.\ Rev.~A \textbf{85},
  021602(R) (2012).

\bibitem{koschorreck2012}
M.~Koschorreck, D.~Pertot, E.~Vogt, B.~Fr\"ohlich, M.~Feld, and M.~K\"ohl,
  Nature (London) \textbf{485}, 619 (2012).

\bibitem{drake2012}
T.~E. Drake, Y. Sagi, R. Paudel, J.~T. Stewart, J.~P. Gaebler, and
  D.~S. Jin, arXiv:1204.0048.

\bibitem{baym1962}
G.~Baym, Phys.\ Rev.\ \textbf{127}, 1391 (1962).

\end{thebibliography}
\end{document}